\documentclass[journal,onecolumn]{IEEEtran}
\usepackage{multirow}
\usepackage{caption}
\usepackage{subfigure}
\usepackage{graphicx}
\usepackage{xcolor}
\usepackage{amsmath}
\usepackage{booktabs}

\newcommand{\specflat}{{spectral\_flatness}}
\newcommand{\snr}{{snr}}

\ifCLASSINFOpdf
\else
\fi

\begin{document}
\title{Houston we have a Divergence: \\ A Subgroup Performance Analysis of ASR Models}


\author{
    \IEEEauthorblockN{Alkis Koudounas, Flavio Giobergia}\\
    \IEEEauthorblockA{Politecnico di Torino
    \\\{firstname.lastname\}@polito.it}
}

\maketitle




\IEEEpeerreviewmaketitle

\section{Abstract}
The Fearless Steps APOLLO Community Resource~\cite{hansen2022fearless} provides an unparalleled opportunity to explore the vast potential of historical multi-speaker team communications from the NASA Apollo missions. This study focuses on discovering which are the characteristics that make Apollo recordings more or less intelligible to Automatic Speech Recognition (ASR) methods. 

We extract, for each audio recording, interpretable metadata on the recordings themselves (signal-to-noise ratio, spectral flatness, presence of pauses, duration of the sentence), on the transcript (number of words spoken, speaking rate) or known \textit{a priori} (speaker). 
Next, we identify subgroups of audio recordings based on combinations of the extracted metadata (e.g., \textit{\{speaker=NEIL, snr=low, total\_duration=low\}} to indicate short, noisy recordings of Neil Armstrong). We compute the performance (e.g., Word Error Rate) for each subgroup and the difference in performance (\textit{``divergence''}) w.r.t the overall population. We rely on the work presented in \cite{pastor2021looking} for the extraction of relevant subgroups and their divergences. 

We then apply the Whisper model~\cite{whisper} in different sizes (\texttt{base}, \texttt{small}, \texttt{medium}, \texttt{large-v3}), trained on English-only or multilingual datasets\footnote{The large model has only been pre-trained on the multilingual dataset}, in zero-shot or after fine-tuning.
We conduct several analyses to (i) automatically identify and describe the most problematic subgroups for a given model, (ii) examine the impact of fine-tuning w.r.t. zero-shot at the subgroup level, (iii) understand the effect of model size on subgroup performance, and (iv) analyze if multilingual models are more sensitive than monolingual to subgroup performance disparities.

\begin{figure}[]
  \centering
    \subfigure[Global Shapely Values for the \texttt{base-en} model. Each bar represents the contribution of the respective metadata to the WER (lower values imply a better WER).] 
    {\centering 
    \includegraphics[width=0.33\textwidth]{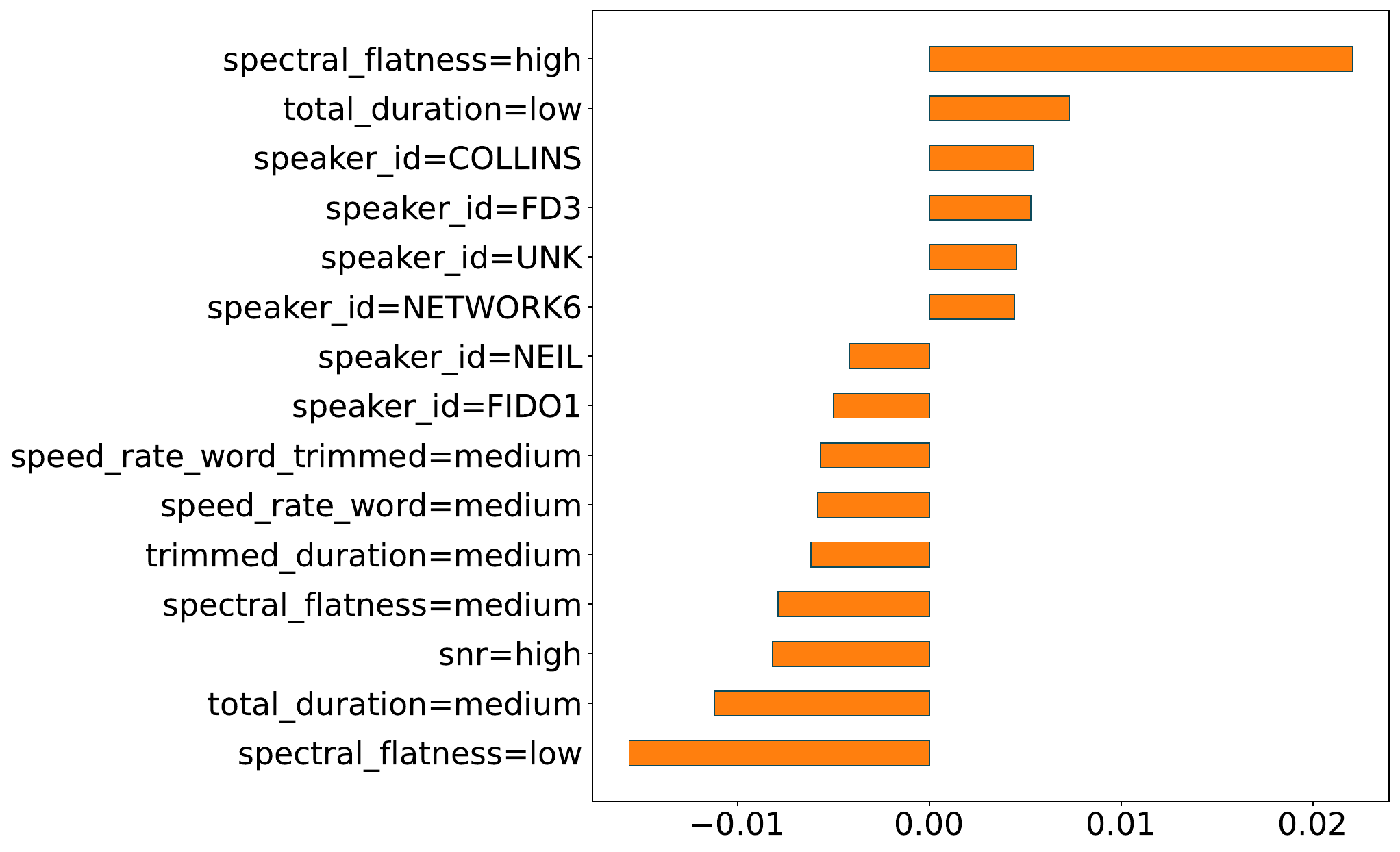}
    \label{fig:gsv}} 
    \hspace{0pt}
    \subfigure[Gain distribution between \texttt{medium-en} and \texttt{medium-multilingual}. In blue, the subgroups where \texttt{medium-en} has an advantage, in orange where \texttt{medium-en} decreases in performance.] 
    {\centering
    \includegraphics[width=0.30\textwidth]{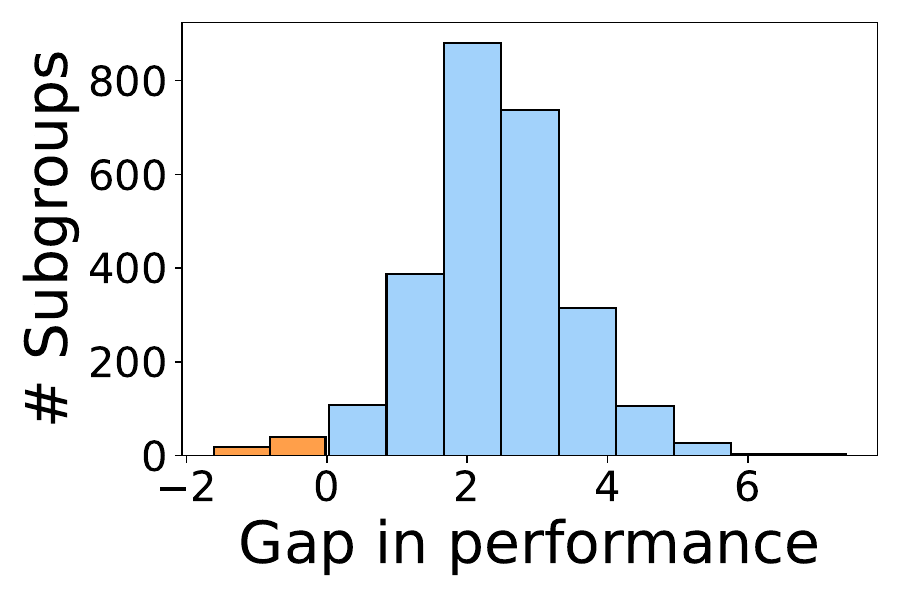}
    \label{fig:distrgain}} 
    \caption{(\textbf{Left}) Global Shapely Values for a single model, (\textbf{Right}) Gain distribution between two models.}
    \label{fig:gsv-distrgain}
\end{figure}

\paragraph{Subgroup Performance Assessment}
We report subgroup-specific performance metrics in Table~\ref{table:subgroups} to identify subgroups with the poorest ASR performance for the \texttt{medium-en} and \texttt{medium-multilingual}  models. This analysis informs us about the challenges posed by different subpopulations within the Fearless Steps APOLLO corpus. We additionally report in Figure \ref{fig:gsv} the Global Shapely Values \cite{pastor2021looking,koudounas2023bad} for the \texttt{base-en} model. This is informative of the type of metadata that affects positively or negatively the predictions. 
For example, recordings with a high SNR and a low spectral flatness are generally associated with better performance (i.e., lower WER), as intuition dictates. Other information about the speakers indicates the extent to which various people are intelligible, according to the ASR model. 

\paragraph{Pre-trained vs. fine-tuned models}
We investigate the impact of fine-tuning on ASR models by comparing the performance of pre-trained models with their fine-tuned counterparts. Table \ref{tab:zsft} shows that fine-tuning consistently globally improves ASR performance. In addition, we show that all divergences get closer to 0: for negative divergences, this means that the fine-tuned model improves on the subgroups that performed poorly in zero-shot). For positive divergences, it means that the fine-tuned model does not improve as much for the already well-performing subgroups. In other words, fine-tuning reduces the divergence of the model across subgroups.

\paragraph{Base vs. large models}
We compare the performance of the base and the small ASR models in a zero-shot setting. It has been shown in literature that larger models do not necessarily consistently outperform smaller models across all subgroups \cite{koudounas2023,koudounas2024taslp}. 
In the first block of Table~\ref{table:comparison}, we analyze the different performance between the \texttt{base-en} and the \texttt{small-en} models. The smaller model outperforms the larger one in some subgroups, even if with a relatively modest gap. Conversely, the improvement observed when transitioning from the base to the small model is substantially higher.

\paragraph{Multilingual vs. English-only Models}
We examine the performance disparity between multilingual and English-only ASR models, identifying subgroups for which the multilingual model consistently outperforms the English-only counterpart. For example, Figure \ref{fig:distrgain} shows the variation in performance between the subgroups, when passing from English-only to multilingual. Interestingly, most subgroups benefit from using the English-only model. However, a small number of subgroups actually decrease in performance. Examples of such subgroups are \{\textit{n\_words=high, snr=high, speaker=FD1, trim\_dur=high}\} and \{\textit{speaker=FD1, spectral\_flatness=low, tot\_dur=high}\}

In conclusion, our study provides a detailed comparative analysis of ASR methods applied to the Fearless Steps APOLLO corpus. The insights gained enhance our understanding of subgroup-specific performance variations, paving the way for advancements in the development and optimization of ASR systems for Earth-to-space communications.

\begin{table*}[!h]
\centering
\setlength\tabcolsep{2pt} 
\scalebox{0.90}
{%
\begin{tabular}{ccccccc}
\toprule

\textit{Model} & \textit{Subgroups} & $WER$ & $\Delta_{WER}$  & \textit{t} \\ \midrule

\multirow{2}{*}{\texttt{medium-en}} 
        & \texttt{S\textsuperscript{-}: }\{\textit{\snr=high, \specflat=medium, speakRate=high, speakRate\_trim=high, trim\_dur=medium}\} & 82.985 & 2.985 & 2.41 \\
        & \texttt{S\textsuperscript{+}: }\{\textit{speaker=CAPCOM1, \specflat=low, speakRate\_trim=high}\} & 76.703 & -3.297 & 2.71 \\ 
\midrule

\multirow{2}{*}{\texttt{medium-multi}} 
        & \texttt{S\textsuperscript{-}: }\{\textit{n\_words=medium, \snr=medium, speaker=FD1}\} & 85.219 & 2.853 & 2.92 \\
        & \texttt{S\textsuperscript{+}: }\{\textit{speaker=BUZZ, \specflat=medium, speakRate\_trim=high}\} & 79.830 & -2.536 & 2.63 \\ 



        
\bottomrule
\end{tabular}
}
\caption{WER divergence for the most negatively (\texttt{S\textsuperscript{-}}) and positively (\texttt{S\textsuperscript{+}}) divergent subgroups compared to overall test performance. The \textit{t} column indicates the Welch's t-test value.}
\label{table:subgroups}
\end{table*}

\begin{table}[!t]
\centering
\scalebox{0.90}
{%
\begin{tabular}{ccclll}
\toprule
\textit{Model} & \textbf{$WER$} & $\Delta_{WER}^{min}$ & $\Delta_{WER}^{max}$ & $\Delta_{WER}^{avg}$ & $\Delta_{WER}^{std}$ \\ \midrule
\texttt{base-en}        & 89.997       & -1.528             & 1.299              &  0.015             & 0.411              \\
\texttt{base-ft}        & 77.444       & -1.077             & 0.983              &  -0.002            & 0.297              \\ \midrule
\texttt{small-en}       & 85.009       & -2.051             & 2.134              &  -0.057            & 0.532              \\
\texttt{small-ft}       & 69.975       & -0.934             & 0.790              &  0.004             & 0.229              \\ \midrule
\texttt{medium-en}      & 80.000       & -3.297             & 2.985              &  -0.050            & 0.797              \\
\texttt{medium-ft}      & 60.028       & -0.637             & 0.883              &  -0.011            & 0.197              \\ \midrule
\texttt{large-v3}       & 75.024       & -1.964             & 1.921              &  0.026             & 0.528              \\
\texttt{large-v3-ft}    & 49.996       & -1.657             & 1.667              &  -0.029            & 0.385              \\ 
\bottomrule
\end{tabular}}
\caption{Information about the Word Error Rate for various models, either fine-tuned or not. Performance reported overall, and in terms of subgroup divergence. For the latter, the spread, minimum, maximum, and average divergences are reported.}
\label{tab:zsft}
\end{table}
\begin{table*}[!t]
\centering
\setlength\tabcolsep{2pt} 
\scalebox{0.85}
{%
\begin{tabular}{cccccccc}
    \toprule

    \textit{Comparison} & \multicolumn{2}{c}{\textit{Subgroups}} & $gap_{WER}$ & $f_\texttt{M1}$ & $f_\texttt{M2}$ & \textit{t} \\ \midrule
    \multirow{2}{*}{\texttt{base-en} vs. \texttt{small-en}} 
        & $\downarrow$  & \{\textit{\snr=high, trim\_dur=low}\}  & 1.04 & 88.76 & 87.80 & 1.99 \\ \cmidrule{2-7} 
        & $\uparrow$ & \{\textit{speaker=MSTC1, trim\_dur=low}\} & -4.67 & 90.71 & 86.05 & 2.21 \\ 
    \midrule

    \multirow{2}{*}{\texttt{medium-en} vs. \texttt{medium-multi}} 
        & $\downarrow$ & \{\textit{\snr=medium, speakRate\_trim=medium, trim\_dur=low}\} & 7.4 & 76.94 & 84.34 & 2.64 \\ \cmidrule{2-7}
        & $\uparrow$ & \{\textit{n\_words=high, \snr=high, spkID=FD1, trim\_dur=high}\} & -1.61 & 82.12 & 80.51 & 2.06 \\ 
    \midrule

    \multirow{1}{*}{\texttt{large-v3} vs. \texttt{large-v3-ft}} 
        & $\uparrow$ & \{\textit{n\_words=low, spkID=FD1, speakRate=medium, speakRate\_trim=medium}\} & -28.09 & 76.94 & 48.86 & 2.30 \\ 
    
    \bottomrule

    \end{tabular}}
\caption{WER Performance gap when changing the model size (\texttt{base} vs. \texttt{small}), the pre-training objective (\texttt{medium-en} vs. \texttt{medium-multi}), or the training methodology (\texttt{zero-shot} or \texttt{fine-tuned}). ($\uparrow$) denotes the highest performance improvement (a lower WER indicates a better score), ($\downarrow$) indicates the largest decrease.}
\label{table:comparison}
\end{table*}

\ifCLASSOPTIONcaptionsoff
  \newpage
\fi

\newpage
\bibliographystyle{IEEEtran}
\bibliography{main}

\end{document}